\documentclass[10pt]{article}

\makeatletter
\newcommand{\xrightleftarrows}[2][]{%
  \ensuremath{%
    \mathrel{%
      \raisebox{-1pt}{%
        \rlap{%
          \raisebox{5pt}{%
            $%
            \ext@arrow 0359\rightarrowupfill@{\hphantom{#1}}{#2}%
            $%
          }%
        }%
        \hbox{%
          $%
          \ext@arrow 3095\leftarrowdownfill@{#1}{\hphantom{#2}}%
          $%
        }%
      }%
    }%
  }%
}
\newcommand*{\leftarrowdownfill@}{%
  \arrowfill@\leftarrow\relbar\relbar
}
\newcommand*{\rightarrowupfill@}{%
  \arrowfill@\relbar\relbar\rightarrow
}
\makeatother
\input
\usepackage{amsmath}
\usepackage{amssymb}

\usepackage{graphicx}

\usepackage{cite}

\usepackage{color} 

\usepackage[labelfont=bf,labelsep=period,justification=raggedright]{caption}

\makeatletter
\renewcommand{\@biblabel}[1]{\quad#1.}
\makeatother

\date{}

\begin{document}

% Title must be 150 characters or less
\begin{flushleft}
{\Large
\textbf{A generic model of dyadic social relationships}
}
% Insert Author names, affiliations and corresponding author email.
\\
Maroussia Favre$^{1,\ast}$, 
Didier Sornette$^{1,2}$ 
\\
\bf{1} Department of Management, Technology and Economics, Swiss Federal Institute of Technology, ETH Z\"urich, Scheuchzerstrasse 7, 8092 Z\"urich, Switzerland
\\
\bf{2} Swiss Finance Institute, c/o University of Geneva, 40 blvd. Du Pont d'Arve, CH-1211 Geneva 4, Switzerland
\\
$\ast$ maroussiafavre@ethz.ch
\end{flushleft}

% Please keep the abstract between 250 and 300 words
\section*{Abstract}
We introduce a model of dyadic social interactions and establish its correspondence with relational models theory (RMT), a theory of human social relationships. RMT posits four elementary models of relationships governing human interactions, singly or in combination: Communal Sharing, Authority Ranking, Equality Matching, and Market Pricing. To these are added the limiting cases of asocial and null interactions, whereby people do not coordinate with reference to any shared principle. Our model is rooted in the observation that each individual in a dyadic interaction can do either the same thing as the other individual, a different thing or nothing at all. To represent these three possibilities, we consider two individuals that can each act in one out of three ways toward the other: perform a social action X or Y, or alternatively do nothing. We demonstrate that the relationships generated by this model aggregate into six exhaustive and disjoint categories. We propose that four of these categories match the four relational models, while the remaining two correspond to the asocial and null interactions defined in RMT. We generalize our results to the presence of N social actions. We infer that the four relational models form an exhaustive set of all possible dyadic relationships based on social coordination. Hence, we contribute to RMT by offering an answer to the question of why there could exist just four relational models. In addition, we discuss how to use our representation to analyze data sets of dyadic social interactions, and how social actions may be valued and matched by the agents.

% Please keep the Author Summary between 150 and 200 words
% Use first person. PLoS ONE authors please skip this step.
% Author Summary not valid for PLoS ONE submissions.
%\section*{Author Summary}

\section*{Introduction} \label{intro}  

In the present work, we are interested in the basic building blocks of social interactions, namely dyadic relationships. Our contribution is to introduce a representation of dyadic relationships that realistically matches an existing theory of human social relationships, relational models theory (RMT) and can be used for theoretical purposes. Moreover, we discuss how to apply our model to computational modeling and analysis.

Our model is based on the fundamental assumption that, in any dyadic interaction, each individual can do either the same thing as the other individual, a different thing, or nothing at all. To represent these three possibilities, it is sufficient to consider that each agent can do X, Y or nothing ($\varnothing$) to the other agent. X and Y are two different ``social actions," in the sense that they intentionally affect their target. Social actions can have positive or negative effects on the receiver's welfare. For example, an agent A could transfer a useful commodity to an agent B, or A could hit and harm B. In what follows, we generally assume that an agent is a person, but it can also represent a social group (e.g. a company, team, nation and so on) that acts as a single entity in specific interactions.

This setting is represented by $A \xrightleftarrows[X/Y/\varnothing]{X/Y/\varnothing} B$. For instance, an interaction in which A does X and B does Y is represented by $A \stackrel{X}{\underset{Y}{\rightleftarrows}} B$. We call the arrows in these symbols ``action fluxes." That model generates a number of possible relationships between the two agents A and B. 

We find that these relationships aggregate into exactly six disjoint categories of action fluxes. These six categories describe all possible relationships arising from our model, singly or in combination. We propose a mapping between these categories and the four basic social relationships, or relational models (RMs), defined by RMT. Namely, four of the six categories map to the RMs, while the remaining two correspond to asocial and null interactions. We argue that this categorization and mapping show that the RMs constitute an exhaustive set of coordinated dyadic social relationships. To take into account that real social interactions involve an infinite variety of social actions, we generalize our model to the presence of any number N of social actions and show that this leads to the same six categories of action fluxes.

Relational models theory was introduced by Alan Fiske \cite{FiskeBook, Fiske} in the field of anthropology to study how people construct their social relationships. RMT posits that people use four elementary models to organize most aspects of most social interactions in all societies. These models are Communal Sharing, Authority Ranking, Equality Matching, and Market Pricing. RMT has motivated a considerable amount of research that supports, develops or applies the theory, not only in its original field of social cognition \cite{Fiske95, Fiskeetal91, Fiske93, FiskeHaslam97}, but also in diverse disciplines such as neuroscience \cite{Iacoboni04}, psychopathology \cite{Haslametal02}, ethnography \cite{Whitehead02}, experimental psychology \cite{Brodbecketal13}, evolutionary social psychology \cite{Haslam97}, and perceptions of justice \cite{Goodnow98}, to name a few. For an overview of this research, see \cite{Haslam04, RMT_overview}. 

\begin{itemize}
\item In the Communal Sharing (CS) model, people perceive in-group members as equivalent and undifferentiated. CS relationships are based on principles of unity, identity, conformity and undifferentiated sharing of resources. Decision-making is achieved through consensus.  CS is typically manifested in close family or friendship bonds, teams, nationalities, ethnicities or between soldiers. 

\item In Authority Ranking (AR) relationships, people are asymmetrically ranked in a linear hierarchy. Subordinates are expected to defer, respect and obey high-rankers, who take precedence. Conversely, superiors protect and lead low-rankers. Subordinates are thus not exploited and also benefit from the relationship. Resources are distributed according to ranks and decision-making follows a top-down chain of command.

\item Equality Matching (EM) relationships are based on a principle of equal balance and one-to-one reciprocity. Salient EM manifestations are turn-taking, democratic voting (one person, one vote), in-kind reciprocity, coin flipping, distribution of equal shares, and tit-for-tat retaliation.

\item The Market Pricing (MP) model is based on a principle of proportionality. Relationships are organized with reference to socially meaningful ratios and rates, such as prices, cost-benefit analyses or time optimization. Rewards and punishment are proportional to merit. Abstract symbols, typically money, are used to represent relative values. MP relationships are not necessarily individualistic; for instance, utilitarian judgments seeking the greatest good for the greatest number are manifestations of MP.
\end{itemize}

The four relational models have in common that they suppose a coordination between individuals with reference to a shared model. To these, Fiske adds two limiting cases that do not involve any other-regarding abilities or coordination \cite{FiskeBook} (pp. 19-20):
\begin{itemize}
\item In asocial interactions, a person exploits others and treats them as animate objects or means to an end (as in psychopathy, armed robbery, pillage);
\item In null ``interactions," people do not interact at all (they do not actively ignore each other, which still requires a coordination), as can be the case of two inhabitants of the same building who never cross each other's way or fail to notice each other's existence, and thus do not adapt their actions to each other.
\end{itemize}

In order to better understand RMT, it is helpful to locate it in the landscape of other social, political and economical theories. Here we follow closely the review made by Senior et al. \cite{Verweij2013} of this theoretical landscape. RMT is identified as a theory of constrained relativism, which lies between the two extremes of rational choice analysis and poststructuralism. Theories belonging to the two latter domains have dominated political science, sociology and economy for several decades, while constrained relativism has had less influence and is not as widely known. Rational choice theory holds that people are fully rational, follow their self-interest and instantly process all available information. Universal analytical models are thus expected to predict the behavior of these rational agents. At the other extreme, poststructuralism posits that every person, society and epoch, is fundamentally unique. According to that view, no generalization can be made; only descriptions are possible and relevant, without offering any prospect of scientific prediction. 

Of the two dominant positions, rational choice theory has been favored in many scientific domains, since it calls for the construction of explanatory and predictive models, forbidden by the very definition of poststructuralism. Yet alternatives to rational choice theory are on the rise, as it is apparent that people are strongly (and primarily) influenced by emotions, feelings and subconscious processes. Notably, rational choice theory fails at explaining or predicting major social, economical or political events, such as financial bubbles and crashes, or social and political revolutions.

Occupying the middle ground between the two extremes of rational choice theory and poststructuralism, theories of constrained relativism are based on the idea that there is a limited number of elementary ways of organizing social relations that serve as building blocks for the infinitely varied aspects of social and political life. Theories of constrained relativism other than RMT are, in particular, the theory of socio-cultural viability (or cultural theory) initiated by Mary Douglas and developed further by others \cite{Douglas78,CulturalTheory90} and Jonathan Haidt's moral foundations theory \cite{Haidt07}, among others.

Until now, the focus of RMT was on people and relationships rather than on abstract representations of the social actions instantiated within relationships. A common point of previous approaches of RMT is that they define the RMs as \textit{cognitive} models. Correspondingly, their implementation has been described in terms of how people mentally represent their relationships, using concepts like group belonging (CS), asymmetrical hierarchies (AR), peer equality (EM) and cost-benefit calculations (MP). Formally, the RMs were compared to the four measuring scales defined by Stevens \cite{Stevens46}: nominal (CS), ordinal (AR), interval (EM) and ratio (MP) \cite{FiskeBook} (pp. 210-223). All of this made any attempt to understand why and how people from widely different cultures manage to coordinate using these same psychological concepts a very ambitious undertaking, and naturally led to consider the evolution and functioning of the human brain, as did Bolender \cite{Bolender} and Iacoboni et al. \cite{Iacoboni04}, for instance. 

Nettle et al. \cite{Nettle} recently opened the way to model what is being transferred from one individual to another. The authors defined three strategies to allocate a resource between two individuals. They presented one of the strategies as typifying CS. Their result was to determine the domain of parameters making each strategy evolutionarily stable. In an analogous modeling spirit, our approach offers an abstract representation of the patterns of social actions performed by dyads in all four relational models, as well as the asocial and null interactions.
 
The rest of this article is organized as follows. In the method, we present our model of action fluxes. In the results, we demonstrate the exhaustiveness of the six categories arising from that representation, and match the categories to the four relational models and the asocial and null interactions. We then generalize the finding of these six categories to the situation involving any number N of social actions. In the discussion, we touch on a method to analyze and interpret data sets of dyadic social interactions. We also express a hypothesis about how social actions are valued and matched by the agents.

% Results and Discussion can be combined.

\section*{Method}

We consider a model of two agents interacting through social actions. A social action corresponds to any action intentionally targeting the receiver and affecting her welfare positively or negatively. It can consist in the transfer of commodities (e.g. objects, food, water, etc.) or services, but can also be a comforting act, talking, harm, violence, and so on. 
 
Let A and B be two distinct agents, and X and Y different social actions. In general, we assume that A and B are two people. However, an agent can also represent a group that acts as a social unit toward a person or another group (e.g. a nation, in the context of its diplomatic relation with another nation). We posit that each agent can act in one out of three ways toward the other agent: do X, Y or nothing ($\varnothing$). The idea at the root of our model is that, in general, each individual in a dyadic interaction can do either the same thing as the other individual, a different thing, or nothing at all. Hence the three possibilities X, Y and $\varnothing$ are sufficient to abstractly describe what can happen in any given interaction between two individuals. Namely, if they do the same thing, they both do X (or Y) ; if they perform different (non-null) actions, one does X and the other does Y. Our setting is represented by $A \xrightleftarrows[X/Y/\varnothing]{X/Y/\varnothing} B$. We call ``action fluxes" the arrows in that symbol. 

The setting $A \xrightleftarrows[X/Y/\varnothing]{X/Y/\varnothing} B$ generates nine cases shown in Table \ref{9poss_table} that we call ``elementary interactions" or just ``interactions." The bottom right case of that table corresponds to the null interaction. For example, the elementary interaction $A \stackrel{X}{\underset{Y}{\rightleftarrows}} B$ means that agent A does X to B, and agent B does Y to A. The elementary interaction $A \stackrel{X}{\underset{\varnothing}{\rightleftarrows}} B$ means that agent A does X to B, without any linked flux going reciprocally from B to A. For convenience of notations, we reduce this symbol to $A\stackrel{X}{\rightarrow} B$. Table \ref{9poss_table2} shows this simplified notation for the interactions with one empty flux (i.e. one null action, $\varnothing$).

\begin{table}[!ht]
\caption{
\bf{Nine elementary interactions}}
\begin{center}
\begin{tabular}{|c|c|c|}
\hline
$A \stackrel{X}{\underset{X}{\rightleftarrows}} B $ & $A \stackrel{X}{\underset{Y}{\rightleftarrows}} B $ & $A \stackrel{X}{\underset{\varnothing}{\rightleftarrows}} B$ \\
\hline
$A \stackrel{Y}{\underset{X}{\rightleftarrows}} B $ & $A \stackrel{Y}{\underset{Y}{\rightleftarrows}} B $ & $A \stackrel{Y}{\underset{\varnothing}{\rightleftarrows}} B$ \\
\hline
$A \stackrel{\varnothing}{\underset{X}{\rightleftarrows}} B$ & $A \stackrel{\varnothing}{\underset{Y}{\rightleftarrows}} B$ & $A \stackrel{\varnothing}{\underset{\varnothing}{\rightleftarrows}} B$ \\
\hline
\end{tabular}
\end{center}
\begin{flushleft}
Each agent (A and B) can do X, Y or $\varnothing$ (nothing) to the other agent. This generates nine possible elementary interactions shown in this table. The bottom right case corresponds to the null interaction.
\end{flushleft}
\label{9poss_table}
\end{table}

\begin{table}[!ht]
\caption{
\bf{Nine elementary interactions, simplified}}
\begin{center}
\begin{tabular}{|c|c|c|}
\hline
$A \stackrel{X}{\underset{X}{\rightleftarrows}} B $ & $A \stackrel{X}{\underset{Y}{\rightleftarrows}} B $ & $A \stackrel{X}{\rightarrow} B$ \\
\hline
$A \stackrel{Y}{\underset{X}{\rightleftarrows}} B $ & $A \stackrel{Y}{\underset{Y}{\rightleftarrows}} B $ & $A \stackrel{Y}{\rightarrow} B$ \\
\hline
$A \stackrel{X}{\leftarrow} B$ & $A \stackrel{Y}{\leftarrow} B$ & $A \stackrel{\varnothing}{\underset{\varnothing}{\rightleftarrows}} B$ \\
\hline
\end{tabular}
\end{center}
\begin{flushleft}
Same as Table \ref{9poss_table}, with simplified notations for the interactions involving one empty flux.
\end{flushleft}
\label{9poss_table2}
\end{table}

We call ``relationship" a realization of one or several elementary interactions between two individuals. A ``composite relationship" is a combination of different elementary interactions, for example [$A\stackrel{X}{\rightarrow} B$ and $A\stackrel{Y}{\leftarrow} B$]. We put between brackets the elementary interactions belonging to a composite relationship, in order to distinguish a composite relationship from a mere enumeration of elementary interactions. A ``simple relationship" corresponds to the occurrence of only one type of elementary interaction, for instance $A \stackrel{X}{\underset{Y}{\rightleftarrows}} B$. 

In both relationships above, A does X and B does Y. We differentiate between these two relationships according to the following rule. We posit that, over time, the (simple) relationship $A \stackrel{X}{\underset{Y}{\rightleftarrows}} B$ entails m fluxes $A\stackrel{X}{\rightarrow} B$ and also m fluxes $A\stackrel{Y}{\leftarrow} B$ in alternation, i.e. each flux $A\stackrel{X}{\rightarrow} B$ is followed by a flux $A\stackrel{Y}{\leftarrow} B$. Every time A does X, B does Y. Correspondingly, if the relationship started with B doing Y, then every time B does Y, A does X. The number m may be equal to 1 (if the interaction occurs just once) or may be larger (if the interaction is repeated). That does not imply that the ``amounts" of X and Y inside the fluxes match, or even that such quantities can be measured. We come back to that possibility in the discussion. Here we are only talking about the number of fluxes, not the weight of their content. For simplicity, we do not specify the number m when we talk about $A \stackrel{X}{\underset{Y}{\rightleftarrows}} B$.

In contrast, we posit that, in the composite relationship [$A\stackrel{X}{\rightarrow} B$ and $A\stackrel{Y}{\leftarrow} B$], fluxes are not alternated, such that there are m fluxes $A\stackrel{X}{\rightarrow} B$ and n fluxes $A\stackrel{Y}{\leftarrow} B$, with $m \neq n$ and $m,n \geq 1$. As time goes by, A does X on m occasions, while B does Y on n occasions, without any pattern of alternation. Again, the quantities of X and Y within the fluxes are not specified. Also, for simplicity, we do not write how many times (m and n) each action flux occurs within the relationship.

For clarity, it is useful to examine how many relationships can be defined in our setting. Each of the nine elementary interactions can be present or not in a relationship. There are thus $2^9=512$ possible relationships. However, by definition, the null interaction $A \stackrel{\varnothing}{\underset{\varnothing}{\rightleftarrows}} B$ cannot coexist with any non-null elementary interaction within the same relationship. Therefore, we really only have eight elementary interactions that can combine to form relationships, giving $2^8=256$ relationships, plus the null relationship that we keep separated. But one of the 256 relationships corresponds to the eight non-null elementary interactions being absent. We identify that relationship with the null relationship. Hence, since we want to count the null relationship only once, our model results in 256 relationships in total. Of these relationships, nine are simple. These are the nine elementary interactions. The other relationships are composite and include between two and eight elementary interactions each. These 256 relationships constitute the ``relationships space" of our model with two social actions.

Our goal is to determine the smallest complete categorization of relationships able to span the relationships space. That is, we want to find ``representative relationships" such that all relationships arising from our model can be expressed in terms of representative relationships, singly or in combination. 

Let us give the example of two individuals in a relationship [$A \stackrel{X}{\underset{X}{\rightleftarrows}} B$ and $A \stackrel{Y}{\underset{Y}{\rightleftarrows}} B$]. They are implementing the same interaction with respect to actions X and Y, respectively. If we posit that $A \stackrel{X}{\underset{X}{\rightleftarrows}} B$ is a representative relationship, we can fully describe A and B's relationship by saying that they are in this representative relationship for both X and Y. 

In what follows, we present the reasoning that leads us to define six representative relationships and demonstrate that these six correspond to an exhaustive categorization of the relationships space.

\section*{Results}

\subsection*{Building six representative relationships}

Based on the fact that two individuals can act in either the same or different ways in an elementary interaction, we categorize the relationships arising from our model by exhausting the distinctions that can be done in that setting: 
\begin{itemize}
\item actions can be null ($\varnothing$) or not (X or Y);
\item agents can perform identical or different actions;
\item in the case of different actions, individuals can be able to exchange roles or not within their relationship. For example, say that A pays B to provide her with goods. This is represented by $A \stackrel{X}{\underset{Y}{\rightleftarrows}} B$, an elementary interaction involving different actions: X for ``giving money" and Y for ``giving goods". Agent A also has the possibility to sell or return goods to B, and this occasionally occurs ($A \stackrel{Y}{\underset{X}{\rightleftarrows}} B$). Overall, this is a relationship written [$A \stackrel{X}{\underset{Y}{\rightleftarrows}} B$ and $A \stackrel{Y}{\underset{X}{\rightleftarrows}} B$] and involving exchangeable roles. Now, say that A pays B so that B protects A ($A \stackrel{X}{\underset{Y}{\rightleftarrows}} B$, with Y representing this time ``protecting"). On the other hand, A is unable to offer any protection to B, so that this never happens nor is expected to happen. This is a relationship consisting of only $A \stackrel{X}{\underset{Y}{\rightleftarrows}} B$ and involving non-exchangeable roles.
\end{itemize}
This leads to six different representative relationships, or six categories of action fluxes, that are summarized in Table \ref{TableBuilding6categories} and explained below. The correspondence between these categories and the RMs is presented later.

\begin{table}[!ht]
\caption{
\bf{Six categories of action fluxes}}
\begin{center}
\begin{tabular}{c|c|c|c|c} 
Category & Fluxes &  Representative & Alternative & RMT \\
& characteristics & relationship & notations & \\
\hline
& Identical actions \\
\hline
1 & Non-null actions& $A \stackrel{X}{\underset{X}{\rightleftarrows}} B$ & $A \stackrel{Y}{\underset{Y}{\rightleftarrows}} B$ & EM \\
\hline
2 & Null actions & $A \stackrel{\varnothing}{\underset{\varnothing}{\rightleftarrows}} B $ & & Null \\
\hline
& Different actions \\
\hline
3  & Non-null actions, & [$A \stackrel{X}{\underset{Y}{\rightleftarrows}} B$ and $A \stackrel{Y}{\underset{X}{\rightleftarrows}} B$] & & MP \\
& exchangeable roles & & & \\
 \hline
4  & Non-null actions, & $A \stackrel{X}{\underset{Y}{\rightleftarrows}} B$ & $A \stackrel{Y}{\underset{X}{\rightleftarrows}} B$ & AR \\
& non-exchangeable roles & & \\
\hline
5 & One null action, & [$A \stackrel{X}{\rightarrow} B$ and $A\stackrel{X}{\leftarrow} B$] & [$A \stackrel{Y}{\rightarrow} B$ and $A\stackrel{Y}{\leftarrow} B$] & CS \\
& exchangeable roles & & & \\
\hline
6  & One null action, & $A \stackrel{X}{\rightarrow} B$  & $A \stackrel{Y}{\rightarrow} B$, or  $A \stackrel{X}{\leftarrow} B$, & Asocial \\
 & non-exchangeable roles & & or $A \stackrel{Y}{\leftarrow} B$ & \\
\hline
\end{tabular}
\end{center}
\begin{flushleft}
Exhaustive categorization of relationships in the model of two agents A and B that can each do X, Y or nothing ($\varnothing$). In elementary interactions, agents can do the same thing or not (i.e. actions can be identical or different) and actions can be null ($\varnothing$) or not (X or Y). Within the relationship, agents can be able to exchange roles or not.
\end{flushleft}
\label{TableBuilding6categories}
\end{table}

\begin{itemize}
\item[1.] Non-null identical actions result in a simple relationship symbolized by $A \stackrel{X}{\underset{X}{\rightleftarrows}} B$, or $A \stackrel{Y}{\underset{Y}{\rightleftarrows}} B$. It does not matter whether the social action is denoted by X or Y. We choose $A \stackrel{X}{\underset{X}{\rightleftarrows}} B$ as the representative relationship in that situation.

\item[2.] Two agents identically doing nothing toward each other are in a null relationship written $A \stackrel{\varnothing}{\underset{\varnothing}{\rightleftarrows}} B$. 

\item[3.] Agents performing different non-null actions and able to exchange roles are in a composite representative relationship [$A \stackrel{X}{\underset{Y}{\rightleftarrows}} B$ and $A \stackrel{Y}{\underset{X}{\rightleftarrows}} B$]. 

\item[4.] In the previous case, if agents cannot exchange roles, the relationship is simple and consists in just $A \stackrel{X}{\underset{Y}{\rightleftarrows}} B$ (or just $A \stackrel{Y}{\underset{X}{\rightleftarrows}} B$: again, the notation used for the actions does not matter). 

\item[5.] When one individual does nothing and the other performs a non-null action in an elementary interaction, and roles are exchangeable in the relationship, it is symbolized by [$A\stackrel{X}{\rightarrow} B$ and $A\stackrel{X}{\leftarrow} B$] (or the same interactions with the notation Y instead of X). 

\item[6.] In the previous case, if roles cannot be exchanged, the relationship consists in only $A\stackrel{X}{\rightarrow} B$ (or the same with Y instead of X, or with the action flux going from B to A). 
\end{itemize}

\subsection*{Proof of exhaustiveness}

Our first result is to prove the proposition that the six categories of action fluxes given in Table \ref{TableBuilding6categories} are exhaustive.

$\mathbf{Proposition} $ $\mathbf{1:}$ To describe all relationships arising from the model $A\xrightleftarrows[X/Y/\varnothing]{X/Y/\varnothing}B$, one needs exactly the six categories of action fluxes defined in Table \ref{TableBuilding6categories}.

\vspace{3mm}
 
\noindent \textit{Proof:} On the one hand, the six categories are mutually disjoint. Indeed, the fluxes characteristics defining the categories do not overlap. For example, two actions cannot be identical and different at the same time. This shows that no less than these six categories could suffice to characterize relationships arising from the setting $A\xrightleftarrows[X/Y/\varnothing]{X/Y/\varnothing}B$.

On the other hand, we noted during the building of the six categories that in some cases, the notation X or Y does not matter, giving rise to alternative notations for some categories. Taking into account these arbitrary choices of notation, the six categories of Table \ref{TableBuilding6categories} cover the nine elementary interactions of Table \ref{9poss_table2}, as is seen by comparing these two tables. Hence, any relationship built on these nine elementary interactions can be expressed in terms of the six categories, singly or in combination. This shows that no more than these six categories are necessary to characterize relationships arising from the setting $A\xrightleftarrows[X/Y/\varnothing]{X/Y/\varnothing}B$. This concludes the proof.

\subsection*{Mapping between the categories of action fluxes and the relational models}

Our second result is to propose a mapping between the six categories of action fluxes defined in Table \ref{TableBuilding6categories} and the four relational models, the asocial and null interactions defined in RMT. The mapping is indicated in the last column of Table \ref{TableBuilding6categories} and is put in words below (in the same order as in the table).

\begin{itemize}

\item[1.] In Equality Matching, actions or items of the same nature are exchanged, usually with a time delay making the exchange relevant. Dinner invitations is a typical example. It is essential in EM that each social action is reciprocated. This is what category 1 captures with the representative relationship $A \stackrel{X}{\underset{X}{\rightleftarrows}} B$.

\item[2.] In the null interaction, people do not interact; this corresponds to empty fluxes in both directions, as in category 2 ($A \stackrel{\varnothing}{\underset{\varnothing}{\rightleftarrows}} B $).

\item[3.] In Market Pricing, one thing is exchanged for another; typically, money or another medium of exchange for a good or service. Agents thus perform different actions in elementary interactions. A defining feature of MP is that a buyer can become a seller and vice-versa toward anybody in a fluid manner, provided that agents possess the right resource or skill. Hence, roles can be exchanged, as in category 3, represented by [$A \stackrel{X}{\underset{Y}{\rightleftarrows}} B$ and $A \stackrel{Y}{\underset{X}{\rightleftarrows}} B$].

\item[4.] As in MP, Authority Ranking relationships involve the exchange of one social action against another. However, AR is not as flexible as MP. To start with, actions are fixed: one of them is typically protection, leadership or management, while the other is obedience, respect, subordination, possibly the payment of a tax under one form or another, and so on. In a well-established relationship, roles are fixed as well: superiors and subordinates may never exchange roles. In social hierarchies mediated by AR, such reversals typically occur infrequently and at the price of spectacular power struggles that cause a period of social and political instability and result in new sets of relationships. These considerations lead us to think of AR as a relationship involving different social actions and non-exchangeable roles, as in category 4, represented by $A \stackrel{X}{\underset{Y}{\rightleftarrows}} B$. We note that the impossibility for agents to exchange roles in category 4 implies that at least one individual does something that the other cannot replicate. It thus has to be something hard to learn or based on innate characteristics (e.g. adult body size), or both; this evokes leadership, dominance, protectiveness, wisdom, experience or popularity. In RMT, these are the typical fundamental determinants of any AR relationship; the non-exchangeability in category 4 connects with the notion of asymmetry present in the RMT description of AR.

\item[5.] In Communal Sharing, people give without counting or expecting a reciprocation, which in our representation translates into the property that each flux going one way does not necessarily entail a reciprocating flux. However, overall, each party contributes to the relationship, such that it is not entirely one-sided. This is represented by category 5 with the relationship [$A \stackrel{X}{\rightarrow} B$ and $A\stackrel{X}{\leftarrow} B$].

\item[6.] In the asocial interaction described by RMT, a person uses others as means, exploits them, or takes from them, possibly by force, whatever can be useful to her. Roles are not exchanged. This corresponds to category 6 where only agent A acts toward B ($A \stackrel{X}{\rightarrow} B$). The asocial interaction of RMT specifically corresponds to the case in which A acts in a harmful, exploitative and abusive way that makes it impossible for B to act back in a socially coordinated way toward A. However, our model generally involves social actions that can be beneficial or harmful to their target. From that point of view, category 6 is more general than the asocial interaction of RMT, and perhaps may best be called ``unilateral," with the asocial interaction being a particular case of that category. 

\end{itemize}

Based on the above mapping, we infer that the categories of action fluxes arising from our model offer suitable abstract representations of the exchange of social actions performed by dyads implementing the RMs. Also, given the exhaustiveness of our categorization, we propose that the four RMs constitute an exhaustive description of coordinated dyadic social relationships. 

Let us now highlight properties of our model and resulting categorization that match important aspects of RMT.

The dyadic property of our model reflects the main focus of RMT. The majority of examples given by Fiske \cite{FiskeBook} from various societies around the world are of interactions between two people and sometimes between two groups. Fiske also uses RMs to characterize groups of more than two individuals in which all members use the same relational model in the context of a social activity. For example, if members of a group are all implementing CS when sharing food, it can be called a ``CS group" with respect to that activity \cite{FiskeBook} (p. 151). Rotating credit associations \cite{FiskeBook} (p. 153) or equal distribution of any common resource are typical examples of EM within a group of more than two people. In such situations of homogeneous collective action, our representation also gives an accurate description of what happens between any two members and thus can be used to characterize the group as a whole.

The six classes of action fluxes that we define are mutually disjoint categories. This is in line with Haslam's proposition \cite{Haslam94} that the relational models are indeed categories, as opposed to ``dimensions" (whereby there would be no well-defined boundaries between the RMs) or ``prototypes" (whereby theoretical, ideal RMs would never be realized by real social interactions; RMs would only be approximated along continuous dimensions). 

Moreover, the disjointness of the categories reflects the view of RMT that any specific aspect of any social interaction corresponds to one, and only one RM (or alternatively, the asocial or the null interaction). This applies to two levels: the way people think of their dyadic relationships with particular persons \cite{Haslam94, Haslam94-2}, and the way people categorize each aspect (e.g. decision making, allocation of resources, organization of work) of the coordination of a particular dyad \cite{Fiske04}.

At the same time, RMT points out that any relationship generally consist in a composite of RMs \cite{FiskeBook} (pp. 155-168). Using Table \ref{TableBuilding6categories}, any composite relationship arising from our model can now be interpreted in terms of the RMs. For example, the relationship [$A \stackrel{X}{\underset{X}{\rightleftarrows}} B$ and $A \stackrel{Y}{\underset{Y}{\rightleftarrows}} B$] is interpreted as EM for both X and Y. The relationship [$A \stackrel{X}{\underset{X}{\rightleftarrows}} B$, $A \stackrel{Y}{\rightarrow} B$ and $A\stackrel{Y}{\leftarrow} B$] is an instance of EM for X and CS for Y. The relationships space also includes cases that are less obvious. For instance, [$A \stackrel{X}{\underset{X}{\rightleftarrows}} B$, $A \stackrel{X}{\underset{Y}{\rightleftarrows}}B$ and $A \stackrel{Y}{\underset{X}{\rightleftarrows}}B$] is interpreted in our categorization as EM for X and MP for X and Y. Yet it is rather odd to imagine two people taking turns at doing X and in parallel trading X for Y. This may correspond to a relationship evolving with time from one RM to the other. The generalization of our model to N social actions, presented in the next section, helps represent any familiar composite relationship.

\subsection*{Generalization to N social actions}

In real social relationships, the number of occurring social actions is expected to be larger than two, which motivates the generalization of our results to any number N of social actions. This is our third result. 

For the generalization that follows, we let X and Y be elements of a larger set $\mathbb{S}$ of N social actions $S_{i}$: $\mathbb{S} = \{ S_{i} | i=1,...,N \}$, such that $X, Y \in \mathbb{S}$, for instance $S_1 \equiv X$ and $S_2 \equiv Y$. 

\vspace{3mm}

\noindent $\mathbf{Proposition}$ $\mathbf{2:}$ In the general case of N non-null social actions ($S_1, S_2, ..., S_N \in \mathbb{S}$, $N \geq 2$), one still needs exactly the six categories of Table \ref{TableBuilding6categories} to describe all possible relationships arising from the setting $A\xrightleftarrows[S_1/S_2/.../S_N/\varnothing]{S_1/S_2/.../S_N/\varnothing}B$.

\vspace{3mm}
 
\noindent \textit{Idea of the proof:} We show that the proof of exhaustiveness of the six categories of Table \ref{TableBuilding6categories} carried out for $N=2$ holds for any $N \geq 2$. Namely, the same process allows to build the same six mutually disjoint categories of action fluxes, and these categories span the relationship space for any $N \geq 2$.

\vspace{3mm}

\noindent \textit{Proof:} In the general case of $N \geq 2$ non-null social actions, there are $2^{N+1}$ elementary interactions and $2^{(N+1)^2-1}$ relationships. 

Cases such as $A\xrightleftarrows[S_3]{S_1+S_2}B$ (where A performs several actions simultaneously) can be written $A\xrightleftarrows[S_3]{S_4}B$ (where $S_4$ is a bundle of actions). More generally, any number of actions can be bundled as in that example. Starting from a set of N social actions, the set $\mathbb{S}$ can include all subsets of that set. (The cardinality of $\mathbb{S}$ is then $2^N$.) Hence, any union of two or more subsets (such as $S_1$ and $S_2$ to give $S_4$) gives another subset that is an element of $\mathbb{S}$.

Then, because there are still two agents and thus at most two different social actions per elementary interaction, the elementary interactions have the same forms as for $N=2$, with additional notations for the social actions.

As an illustration, Table \ref{16possN3} shows the sixteen elementary interactions that result from the case $N=3$, i.e. the model $A\xrightleftarrows[X/Y/Z/\varnothing]{X/Y/Z/\varnothing}B$. 

\begin{table}[!ht]
\caption{
\bf{Sixteen elementary interactions for $N=3$ social actions}}
\begin{center}
\begin{tabular}{|c|c|c|c|}
\hline
$A \stackrel{X}{\underset{X}{\rightleftarrows}} B$ & $A \stackrel{X}{\underset{Y}{\rightleftarrows}} B $ & $A \stackrel{X}{\underset{Z}{\rightleftarrows}} B $ & $A \stackrel{X}{\rightarrow} B$ \\
\hline
$A \stackrel{Y}{\underset{X}{\rightleftarrows}} B $ & $A \stackrel{Y}{\underset{Y}{\rightleftarrows}} B $ & $A \stackrel{Y}{\underset{Z}{\rightleftarrows}} B $ & $A \stackrel{Y}{\rightarrow} B$ \\
\hline
$A \stackrel{Z}{\underset{X}{\rightleftarrows}} B$ & $A \stackrel{Z}{\underset{Y}{\rightleftarrows}} B$ & $A \stackrel{Z}{\underset{Z}{\rightleftarrows}} B$ & $A \stackrel{Z}{\rightarrow} B$ \\
\hline
$A \stackrel{X}{\leftarrow} B$ & $A \stackrel{Y}{\leftarrow} B$ & $A \stackrel{Z}{\leftarrow} B$ & $A \stackrel{\varnothing}{\underset{\varnothing}{\rightleftarrows}} B$ \\
\hline
\end{tabular}
\end{center}
\begin{flushleft}
This table shows the sixteen elementary interactions arising from our model with $N=3$ non-null social actions X,Y,Z between two agents A and B, that is, $A\xrightleftarrows[X/Y/Z/\varnothing]{X/Y/Z/\varnothing}B$. We use simplified notations for the interactions involving one empty flux.
\end{flushleft}
\label{16possN3}
\end{table}

For any $N \geq 2$, looking at an elementary interaction between two individuals, one can still only differentiate between (i) identical or different actions, (ii) interchangeable or non-interchangeable roles, (iii) null or non-null actions. Hence, with more than two actions, this differentiation process leads to the same six disjoint categories, except with more alternative notations than in Table \ref{TableBuilding6categories}.

For example, for $N=3$, category 1 (EM) gets one more alternative notation than for $N=2$, namely $A \stackrel{Z}{\underset{Z}{\rightleftarrows}} B$. Category 3 (MP) gets two alternative notations: [$A \stackrel{X}{\underset{Z}{\rightleftarrows}} B$ and $A \stackrel{Z}{\underset{X}{\rightleftarrows}} B$], and [$A \stackrel{Y}{\underset{Z}{\rightleftarrows}} B$ and $A \stackrel{Z}{\underset{Y}{\rightleftarrows}} B$]. Category 4 (AR) gets four more alternative notations: $A \stackrel{X}{\underset{Z}{\rightleftarrows}} B$, $A \stackrel{Z}{\underset{X}{\rightleftarrows}} B$, $A \stackrel{Y}{\underset{Z}{\rightleftarrows}} B$, and $A \stackrel{Z}{\underset{Y}{\rightleftarrows}} B$. Category 5 (CS) gets one more alternative notation, [$A \stackrel{Z}{\rightarrow} B$ and $A\stackrel{Z}{\leftarrow} B$]. Finally, category 6 (asocial) gets two more alternative notations, $A \stackrel{Z}{\rightarrow} B$ and $A \stackrel{Z}{\leftarrow} B$.

For any $N \geq 2$, all of the $2^{N+1}$ elementary interactions are included in the representative relationships of the six categories and their alternative notations. This results from the building process of the six categories, with the differentiations covering all possible cases.

In the example of $N=3$, the above statement is illustrated by the comparison of the sixteen elementary interactions of Table \ref{16possN3} with the six categories and their alternative notations for $N=2$ (Table \ref{TableBuilding6categories}), completed by the alternative notations for $N=3$ listed above.

This concludes the proof of exhaustiveness of the six categories of Table \ref{TableBuilding6categories} for any number $N \geq 2$ of non-null social actions. 

With N social actions at hand, richer composite relationships can be represented. Let us translate into our action fluxes representation an example of composite relationship given by Goldman \cite{Goldman93} (pp. 344-345). Namely, ``two friends may share tapes and records freely with each other (CS), work on a task at which one is an expert and imperiously directs the other (AR), divide equally the cost of gas on a trip (EM), and transfer a bicycle from one to the other for a market-value price (MP)." This gives [$A \stackrel{S_1}{\rightarrow} B$, $A \stackrel{S_1}{\leftarrow} B$, $A \stackrel{S_2}{\underset{S_3}{\rightleftarrows}} B$, $A \stackrel{S_4}{\underset{S_4}{\rightleftarrows}} B$, $A \stackrel{S_5}{\underset{S_6}{\rightleftarrows}} B$, $A \stackrel{S_6}{\underset{S_5}{\rightleftarrows}} B$]. Here the relationship was known and we wrote it in terms of action fluxes. The next step is to find out how to identify a relationship when the action fluxes are given. We touch on how to achieve this in the discussion.

\section*{Discussion}

\subsection*{Analyzing data sets}

Our representation in action fluxes provides a tool to identify types of dyadic relationships occurring within potentially large data sets of social interactions. Both collective and dyadic interactions may occur in real social contexts, but our approach applies specifically to the latter. 

Large data sets can result from any type of online social network or massively multiplayer online role-playing games (MMORPG), for instance. MMORPGs bring hundreds of thousands of players together to cooperate and compete by forming alliances, trading, fighting, and so on, all the while recording every single action and communication of the players. They are used in quantitative social science, for example by Thurner in the context of the game Pardus \cite{Thurner2010Multirel,Thurner2012Zipf, ThurnerSornette2014}. Ethnological and anthropological studies can provide rich reports of social interactions occurring in non-artificial settings that could be coded and interpreted with the aid of our categorization. Data sets of dyadic interactions can also be generated by computer simulations such as agent-based models (ABMs) to test specific questions.

We offer the sketch of a method to analyze a potentially large data set of dyadic social interactions expressed as action fluxes (``A does X to B", etc.). Given a data set involving a number of individuals, one needs to consider separately each pair of individuals. For each pair, one shall examine each social action and test into which category of action fluxes it falls, possibly jointly with another social action (in the case of MP and AR). In its second column, Table \ref{fluxes_table} specifies the patterns of fluxes expected to be observed in each category. 

\begin{table}[!ht]
\caption{
\bf{Detection of categories of action fluxes in data sets of dyadic interactions
}}
\begin{center}
\begin{tabular}{c|c|c|c} 
Category & Pattern of observed fluxes & Representative & RMT \\
& & relationship & \\
\hline
1 & Alternated fluxes $A \stackrel{X}{\rightarrow} B$ and $A \stackrel{X}{\leftarrow} B$ & $A \stackrel{X}{\underset{X}{\rightleftarrows}} B $ & EM \\
\hline
2  & No fluxes between A and B & $A \stackrel{\varnothing}{\underset{\varnothing}{\rightleftarrows}} B$ & Null \\
\hline
3  & Alternated fluxes $A \stackrel{X}{\rightarrow} B$ and $A \stackrel{Y}{\leftarrow} B$, & [$A \stackrel{X}{\underset{Y}{\rightleftarrows}} B$  & MP \\
& and separately, & and & \\
& alternated fluxes $A \stackrel{Y}{\rightarrow} B$ and $A \stackrel{X}{\leftarrow} B$ & $A \stackrel{Y}{\underset{X}{\rightleftarrows}} B$] & \\
\hline
4  & Alternated fluxes $A \stackrel{X}{\rightarrow} B$ and $A \stackrel{Y}{\leftarrow} B$ & $A \stackrel{X}{\underset{Y}{\rightleftarrows}} B$ & AR \\
\hline
5 & Fluxes $A \stackrel{X}{\rightarrow} B$ and $A \stackrel{X}{\leftarrow} B$, & [$A \stackrel{X}{\rightarrow} B$ and & CS \\
& not systematically alternated & $A \stackrel{X}{\leftarrow} B$] &  \\
\hline
6  & Fluxes $A \stackrel{X}{\rightarrow} B$ & $A \stackrel{X}{\rightarrow} B$ & Asocial \\ 
\hline
\end{tabular}
\end{center}
\begin{flushleft}
Patterns of action fluxes expected to be observed in each category. X and Y are social actions belonging to a set $\mathbb{S}$ of size N. A and B are agents.
\end{flushleft}
\label{fluxes_table}
\end{table}

Let us stress the following points:
\begin{itemize}
\item The patterns of observed fluxes given in Table \ref{fluxes_table} are not meant as definitions of the categories. They rather correspond to properties of the fluxes in each category.
\item Linked to the previous point, we stress that the patterns given in that table are not mutually disjoint. They should thus be tested in a certain order. MP should be tested before AR and CS, and category 6 (asocial) should be tested after all other categories. Let us illustrate that point with an example: if A and B are in an MP relationship, one observes alternated fluxes $A \stackrel{X}{\rightarrow} B$ and $A \stackrel{Y}{\leftarrow} B$, as well as alternated fluxes $A \stackrel{Y}{\rightarrow} B$ and $A \stackrel{X}{\leftarrow} B$, re-written as [$A \stackrel{X}{\underset{Y}{\rightleftarrows}} B$ and $A \stackrel{Y}{\underset{X}{\rightleftarrows}} B$]. Yet, A and B's relationship will also respond positive to a CS test, because non-alternated fluxes $A \stackrel{X}{\rightarrow} B$ and $A \stackrel{X}{\leftarrow} B$ are present overall in the relationship (indicative of CS for X), as well as non-alternated fluxes $A \stackrel{Y}{\rightarrow} B$ and $A \stackrel{Y}{\leftarrow} B$ (CS for Y). In that case, the alternation that marks an MP relationship should prevail in the observer's interpretation because it is very unlikely to happen by chance alone.
\item A tolerance level should be defined for the alternation of fluxes (in EM, MP and AR). For instance, the alternation of fluxes in an EM relationship does not need to be strict. In real conditions, people can be flexible and take several turns in a row before the other party reciprocates. Occasional cheating or inexact record keeping can also occur within an otherwise stable relationship. Hence, a low tolerance could lead to false negatives, whereby the observer would miss situations of EM, MP and AR. On the other hand, high tolerance levels might lead to wrong interpretations: fluxes could be falsely interpreted as manifestations of EM, MP or AR. The adequate tolerance level may vary per data set or relationship and may be checked against the individuals' communications, if available.
\item Relationships may change over time. Individuals may initiate a certain relationship that transforms over time into another, linked to increased trust (or mistrust), availability of resources in the environment, and so on. To detect such changes, analyses can be carried out over different time windows.
\item In any real data set, individuals may belong to larger social units that interact with each other. If blindly applied to all pairs of individuals, our model may miss these high-level effects. For instance, say that agent A from group $G_1$ attacked B from group $G_2$. Agent C from $G_2$, feeling very close to B (perhaps in a CS way), decides to punish A and attacks her in an eye-for-an-eye, tooth-for-a-tooth fashion (EM). If one knows nothing of the groups existence, one analyzes separately the pairs (A, B) and (A, C). From the observations $A \stackrel{X}{\rightarrow} B$ and $A \stackrel{X}{\leftarrow} C$, one concludes to the presence of an asocial interaction (category 6) between A and B, as well as between A and C. If however one knows of the groups and applies our model to the high-level agents $G_1$ and $G_2$, one observes $G_1 \stackrel{X}{\underset{X}{\rightleftarrows}} G_2$ and interprets it as an EM relationship between the two groups. This example motivates to identify the relevant social units in the data set under study. This may be achieved by measuring the number and/or duration of interactions between all pairs of individuals and thus creating a weighted graph. By choosing a threshold for the weight of the links, one may be able to isolate social units. Our model would then apply to pairs of members of the same social unit, as well as to pairs of social units. 
\end{itemize}

\subsection*{Valuing the action fluxes}

Our fluxes representation reflects only the presence or absence of fluxes, without saying anything about the quantities involved. The amount carried by an action flux can be readily measured when the action consists in the transfer of a physical item for which a unit of measure can easily be agreed upon. Valuing an action in general is not straightforward. The average time or physical effort necessary to perform the action can be measured (or simply intuitively approximated), but it is naturally much harder to quantitatively agree on a possible emotional or intellectual value. We propose that a value function does not need to be identical for each agent, and each agent does not need to possess a fixed, deterministic value function. For our present needs, it is sufficient for each agent to have personal notions of the value of the social actions performed by herself and others. These personal scales may be probabilistic, in the sense that the value they return may follow probability distributions. Correspondingly, decisions may be probabilistic, as suggested by Quantum Decision Theory \cite{QDT10, QDT11}. Alternatively, a value function could be a von Neumann-Morgenstern utility \cite{NeumannMorgenstern53}, a subjective cumulative prospect utility \cite{KahnemanTversky79}, or any other value function capturing different forms of happiness or contentment, as in the theory of utilitarianism \cite{Mill06}.

We hypothesize that, in a population of interacting agents evolving under selective pressure, the action fluxes of our model converge toward equilibria characterized by an equality in value between opposite fluxes. This proposition rests on the idea that unequal fluxes disadvantage at least one party. They are thus likely to jeopardize the relationship in the short term (in case of inequity aversion), and hinder the survival or reproductive success of the disadvantaged party in the long run. Hence, both individual optimization and selection pressure from external forces in the environment should drive interactions toward stable equilibria characterized by value equalities. We expect these equilibria to depend upon initial conditions and previous states. In other words, different societies would estimate that different things or actions have equal values.

Let us now examine this suggestion in relation to RMT. MP requires a formal matching agreement stating the respective values of what is exchanged, whether actions (such as work), commodities or symbolic items (e.g. money). In EM, the things exchanged are not only of the same nature, but also of the same value. Thus, the idea of value equalities is already embedded into the definitions of these two RMs. RMT keeps CS and AR apart by stating that these RMs are not supposed to necessitate any kind of counting. It may be the case that CS relationships are established only between individuals so close that individual optimization does not occur, such that these relationships may not need to rest on equal contributions overall. 

For its part, the RMT definition of AR is based on the presence of a linear hierarchy and states that superiors generally get more and better things, but have the obligation to act generously according to the principle ``noblesse oblige" \cite{FiskeBook} (pp. 42-43). There is a deep principle of asymmetry and inequality, expressed for instance in \cite{Goldman93} (pp. 343-344): ``When people transfer things from person to person in an AR mode, higher-ranking people get more and better things, and get them sooner, than their subordinates. Higher-ranking people may preempt rare or valuable items, so that inferior people get none at all." In that quote, only one side of the relationship is looked at, namely what the higher-ranking people get from subordinates. Yet AR relationships entail an exchange of protection (or management, etc.) in return for obedience, loyalty, tax payments, and so on. In our representation, the equality would be between the protection offered by the leader and the obedience of the subordinates, whereby the leader may well get ``more and better things," but matching in value the safety she offers to her subjects. Nevertheless, it may be that respective contributions match only in idealized AR relationships, because in practice it is difficult for subordinates to monitor and enforce equality in an essentially asymmetrical relationship.

Another point concerning AR is that, according to Fiske \cite{FiskeBook} (p. 209), the distance between ranks is not socially meaningful; only the linear ordering of ranks is (i.e. which rank is higher, without specifying how much higher). Yet, a value function would allow to measure the distance between ranks. We point out that just because a value function is introduced does not mean that the use of AR requires any computation from the agents. Just as we adapt our every move to the law of gravity without solving mentally at each instant the corresponding equations, or as dogs catch frisbees using simple heuristics \cite{DogsFrisbees,Gigerenzer2000}, it is perfectly conceivable that we are able to recognize and interact with individuals of different ranks without using or having defined any measure of ranks differences or action values. In the case of humans, these heuristics are facilitated by evolved language and culture, which permit the existence of predefined roles (for instance ``chief" or ``servitor") offering an idea of what is expected from each party.

An agent-based model would be a convenient approach to observe and test the evolution of a system toward value equalities. Naturally, it would also be of high interest to examine real social relationships in the making. This, however, raises practical difficulties such as the fact that even new relationships develop within a cultural context that largely predefines how RMs should be implemented, making transient forms unlikely to occur or last long enough to be observed.

% You may title this section "Methods" or "Models". 
% "Models" is not a valid title for PLoS ONE authors. However, PLoS ONE
% authors may use "Analysis" 
%\section*{Materials and Methods}

\section*{Conclusion}

We introduced a model of social interactions between a pair of individuals A and B, each of whom can perform a social action X, Y or nothing, symbolized by $A \xrightleftarrows[X/Y/\varnothing]{X/Y/\varnothing} B$. We demonstrated that from this setting arise six exhaustive and disjoint categories of relationships, four of which match the relational models of RMT, while the remaining two are identified as the asocial and the null interactions. We generalized this categorization to the case of any number N of social actions. We proposed that the categories of action fluxes offer suitable abstract representations of the social actions performed by dyads implementing the relational models. Hence, simulated or real dyads may exhibit various patterns of interactions that can be matched to the six categories of action fluxes, singly or in combination. In that spirit, we discussed a method to identify relational models, expressed as categories of action fluxes, in data sets of dyadic interactions. Finally, we expressed a hypothesis about how social actions are valued and matched by the agents. Our representation can be used to interpret social relationships in terms of RMT and test various hypotheses on dyadic social interactions occurring in potentially large data sets.

% Do NOT remove this, even if you are not including acknowledgments
\section*{Acknowledgments}

We are grateful to Alan Fiske for numerous discussions both at UCLA and Z\"urich, which helped shape and refine our understanding of RMT. We thank Alexandru Babeanu for his thorough review of our model and Marco Verweij for stimulating debates. Finally, we thank the Relational Models Theory International Lab, Christine Sadeghi and Ryan Woodard for insightful feedbacks. %This work has been partially supported by the Swiss National Foundation. TODO plos one wants this in their online system, not in the paper.

%\section*{References}
% The bibtex filename
%\bibliography{plos_template}  

%\section*{Figure Legends}
%\begin{figure}[!ht]
%\begin{center}
%%\includegraphics[width=4in]{figure_name.2.eps}
%\end{center}
%\caption{
%{\bf Bold the first sentence.}  Rest of figure 2  caption. Caption 
%should be left justified, as specified by the options to the caption 
%package.
%}
%\label{Figure_label}
%\end{figure}

%\section*{Tables}

\end{document}